\documentclass[10pt,aps,prl,twocolumn,superscriptaddress,nobibnotes,nodoi,reprint,longbibliography]{revtex4-2}

\usepackage{graphicx}
\usepackage{amsmath,physics,bm,float}
\usepackage{xcolor}
\usepackage{soul}
\usepackage[colorlinks=true,citecolor=blue,urlcolor=blue,linkcolor=blue]{hyperref}
\usepackage{amssymb}
\usepackage{bbm}
\usepackage{natbib}
\usepackage{upgreek}
\usepackage{mathtools}
\usepackage{siunitx}
\usepackage[normalem]{ulem}

\def\dd{{\rm d}}

\DeclareGraphicsExtensions{.pdf,.png,.jpg}
\graphicspath{{./Pictures/}}

\begin{document}
\title{Jerky Motion of Active Granular Particles}
    
	\author{Alexander P.\ Antonov}
	\email{alexander.antonov@hhu.de}
	\affiliation{
		Institut f{\"u}r Theoretische Physik II: Weiche Materie,
		Heinrich-Heine-Universit{\"a}t D{\"u}sseldorf, Universit{\"a}tsstra{\ss}e 1,
		D-40225 D{\"u}sseldorf, 
		Germany}

    \author{Marco Musacchio}
	\affiliation{
		Institut f{\"u}r Theoretische Physik II: Weiche Materie,
		Heinrich-Heine-Universit{\"a}t D{\"u}sseldorf, Universit{\"a}tsstra{\ss}e 1,
		D-40225 D{\"u}sseldorf, 
		Germany}

	\author{Hartmut L{\"o}wen}
	\affiliation{
		Institut f{\"u}r Theoretische Physik II: Weiche Materie,
		Heinrich-Heine-Universit{\"a}t D{\"u}sseldorf, Universit{\"a}tsstra{\ss}e 1,
		D-40225 D{\"u}sseldorf, 
		Germany}

    \author{Lorenzo Caprini}
    \email{lorenzo.caprini@uniroma1.it}
	\affiliation{
		Physics Department, University of Rome La Sapienza, P.le Aldo Moro 5, IT-00185 Rome, Italy}

\date{\today}
\begin{abstract}

        Abrupt transitions between rest and motion can render the standard Newtonian description -- based on position, velocity, and acceleration -- incomplete, requiring higher-order derivatives such as the jerk, the third time derivative of position. Here, we show that the interplay between activity and dry friction gives rise to robust jerk-dominated dynamics in self-propelled particles: the particle speed increases quadratically with time under an active force, in contrast to the linear growth expected for conventional Newtonian dynamics. We demonstrate this behavior analytically, numerically, and experimentally using active vibrobots self-propelling on a vertically vibrating plate at low vibration amplitudes, where surface asperities generate dry friction and, thus, give rise to jerky motion when combined with activity.
        Our results establish dry friction as a simple mechanism for realizing higher-order dynamics in active matter and suggest that jerky dynamics may arise broadly in nonequilibrium systems with frictional contacts.
       
\end{abstract}

\maketitle

{\paragraph{Introduction}--}
Modern classical mechanics is rooted in Newtonian mechanics~\cite{newton1833philosophiae}, which is widely regarded as its earliest formulation. Though famously associated with three laws of motion, Newtonian mechanics also involves several additional postulates, including inertia (mass) positivity, mass additivity, and superposition principle~\cite{greiner2004classical}. Together, they imply that the kinematics of an object is completely described by its position, velocity, and acceleration. However, in systems exhibiting abrupt transitions between rest and motion, these variables may not fully capture the underlying dynamics. In such cases, the equation of motion may involve \textit{jerk} -- the third time derivative of position, or equivalently, the rate of change of acceleration -- with higher-order derivatives potentially entering the description as well~\cite{eager2016beyond}. Notable examples of jerky dynamics arise in a wide range of physical and engineering systems, including chaotic nonlinear dynamical systems~\cite{sprott1997some, von1998all, linz2000no, patidar2005bifurcation, njitacke2017antimonotonicity, li2023some}, traffic flow control~\cite{liu2016tdgl, cheng2017new, zhai2018new}, the Abraham-Lorentz force describing radiation reaction~\cite{Johnson/Hu:2002, Linz/etal:2014}, and cosmological models~\cite{poplawski2006cosmic, tiwari2022anisotropic}. While jerk can arise whenever the forces acting on a system vary along its trajectory, systems out-of-equilibrium provide a particularly promising setting where abrupt changes in the driving dynamics can take place.

Active matter~\cite{marchetti2013hydrodynamics, bechinger2016active, te2026colloquium} represents a prominent example of such systems, where a continuous influx of energy, balanced by environmental dissipation, gives rise to self-propelled motion and drives the system away from equilibrium. Moreover, active matter can locally violate the action-reaction law through non-reciprocal interactions~\cite{saha2020scalar,dinelli2023non, gompper20252025,loos2020irreversibility,fruchart2026nonreciprocal}, providing a natural framework to explore dynamics beyond the Newtonian paradigm. The interplay between energy injection and dissipation can generate variations in acceleration~\cite{fily2014dynamics, mallory2018active}, suggesting that active matter may exhibit jerky motion. However, experimentally observing jerk-dominated dynamics in active matter remains challenging because many active systems operate in the overdamped regime~\cite{bechinger2016active, elgeti2015physics}, where even second-order temporal derivatives are suppressed, or are subject to inertial dynamics in which translational or rotational inertia is significant, such as active granular particles~\cite{scholz2018inertial,caprini2026active}. By contrast, systems exhibiting pronounced third or higher-order dynamics typically rely on additional mechanisms, such as memory \cite{te2021jerky} or time-delayed feedback~\cite{lowen2025gigantic,jose2025jerky}.

\begin{figure*}[ht!]
	\includegraphics[width=\linewidth]{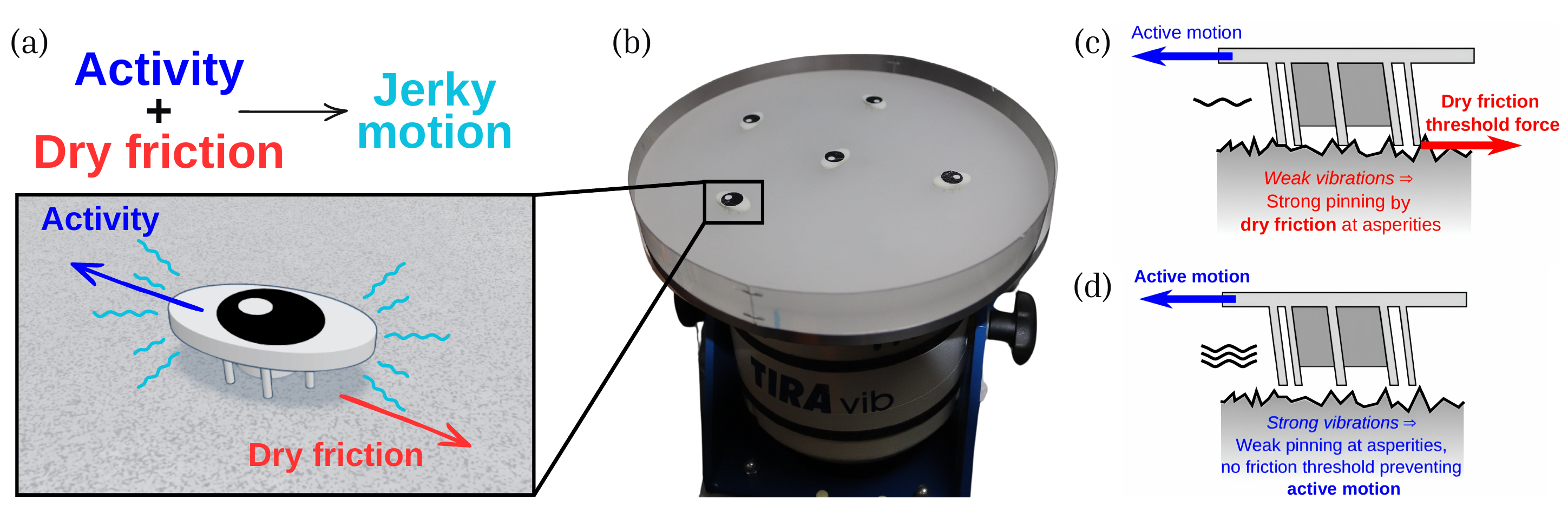}
     \caption{(a) Illustration of an active particle subject to dry friction, giving rise to jerky motion.
     (b) Experimental setup consisting of active vibrobots on a vertically vibrating plate.
     (c,d) Schematic illustration of the frictional contact. Under weak plate vibrations, the vibrobot remains strongly pinned at surface asperities, preventing active motion until the active force exceeds the dry-friction threshold (c). In this regime, the interplay between activity and dry friction gives rise to jerky motion. Under strong vibrations, pinning is greatly reduced (d), rendering dry friction ineffective and allowing persistent active motion.}
    \label{fig:1}
	\end{figure*}

Here, we discover that the interplay between self-propulsion and dry friction gives rise to jerk-dominated active dynamics (Fig.~\ref{fig:1}(a)). Dry friction is characterized by a threshold force: unless the applied force exceeds this threshold, motion cannot occur. We show that the presence of such threshold suppresses lower-order dynamical responses and promotes a trackable jerky motion, providing a simple mechanism for the emergence of higher-order dynamics without requiring memory or time-delayed feedback. We validate this prediction experimentally using active vibrobots (Fig.~\ref{fig:1}(b)) in a weakly vibrated regime, where they exhibit self-propelled motion governed by dry friction arising from contact-induced pinning at surface asperities.

{\paragraph{Jerky motion in active vibrobots} --}
In contrast to conventional Newtonian dynamics, which is governed by second-order equations of motion, jerky motion is described by a third-order differential equation for the particle position $\mathbf{r}$ \cite{linz1998newtonian},
\begin{equation}
\lambda \dddot{\mathbf{r}} = \mathbf{F}(\mathbf{r},\dot{\mathbf{r}},\ddot{\mathbf{r}}),
\label{eq:jerk}
\end{equation}
where $\lambda$ is the jerk coefficient and $\mathbf{F}$ is a generalized force that depends on the particle position, velocity $\mathbf{v}=\dot{\mathbf{r}}$, and acceleration, $\dot{\mathbf{v}}=\ddot{\mathbf{r}}$. For a particle initially at rest and subjected to a constant force, Eq.~\eqref{eq:jerk} predicts a speed that increases quadratically with time,
\begin{equation}
\label{eq:v_scaling_time}
v(t)=|\dot{\mathbf{r}}(t)|\sim t^2,
\end{equation}
and a displacement $\delta r(t)\equiv |\mathbf{r}(t)-\mathbf{r}(0)|$ that scales as $\delta r(t)\sim t^3$. By contrast, under the same conditions, conventional Newtonian dynamics predicts $v(t)=|\dot{\mathbf{r}}(t)|\sim t$ and a corresponding displacement $\delta r(t)\sim t^2$.

	 \begin{figure*}[ht!]
		\includegraphics[width=\linewidth]{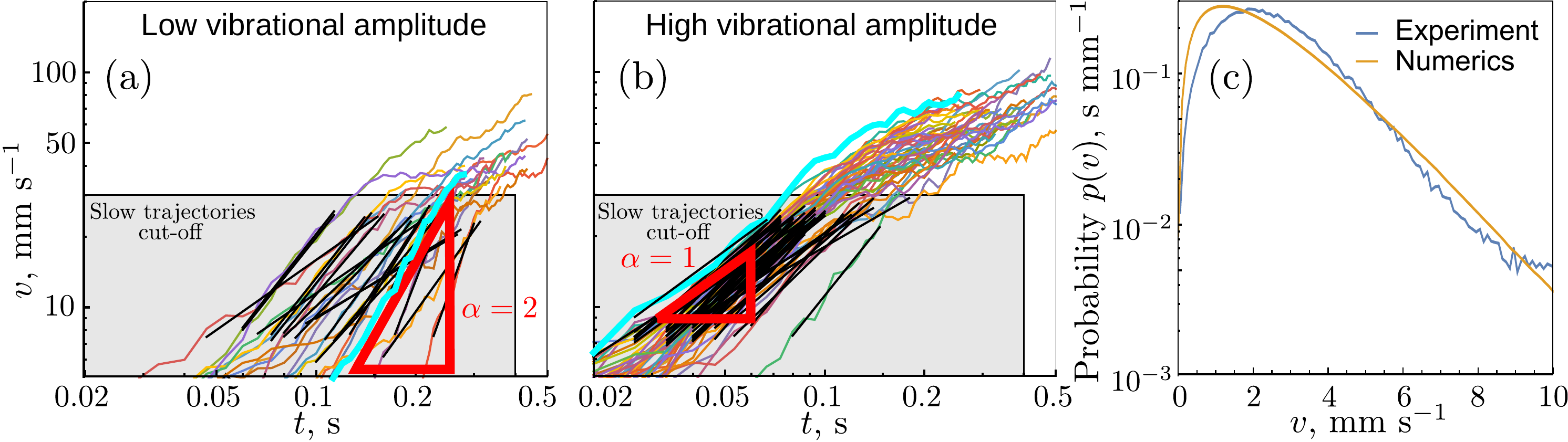}
     \caption{(a-b) Speed evolution analysis of experimental vibrobots initiating their motion after being at rest for (a) a regime dominated by dry friction (shaker's frequency \SI{75}{\hertz} and amplitude $51.186 \pm 0.160$ \si{\micro\meter}, 20 trajectories), and (b) pinning-free regime (shaker's frequency \SI{75}{\hertz}, amplitude $65.200 \pm 0.217$ \si{\micro\meter}, 43 trajectories). In the experiment, slow particles are excluded by imposing a threshold speed that must be achieved within a prescribed time window, as indicated by the gray rectangle. These results reveal much sharper speed increase for the dry friction case. The fitted exponent of $v \sim t^\alpha$, averaged over $v\in[7,25]$, yields (a) $\alpha = 1.84 \pm 0.18$, consistent with jerky motion within uncertainty, and (b) $\alpha = 1.28 \pm 0.03$, approaching the accelerated-motion regime. Representative trajectories for both motion types are highlighted in cyan, and red triangles serve as a guide for an eye. (c) Speed probability distribution $p(v),\, v\equiv |\mathbf{v}|$. The blue line represents experimental data, with orange line corresponding to simulations of Eqs.~\eqref{eq:EOM} with parameters: $m = \SI{1.45}{\gram}$, $\tau = \SI{1}{\second}$, $K = \SI{0.9e-6}{\kilogram\squared\second\per\meter\squared}$, $f_A = \Delta_F =\SI{0.7}{\milli\newton}$.
     }
    \label{fig:2}
	\end{figure*}

Here, we experimentally discover that jerky motion can be generated by the interplay between activity and dry friction.  Our experiment consists of an active vibrobot placed on a vibrating plate \cite{kudrolli2010concentration,aranson2007swirling,kumar2014flocking,carrillo2025depinning,caprini2026active} driven by an electromagnetic shaker (Fig.~\ref{fig:1}(a)-(b)). The vibrobot is a 3D-printed object that contacts the substrate through seven legs (see End Matter). The legs are tilted in the same direction, which leads to active motion of the vibrobot: upon collision of the legs with the rough surface, vibrational kinetic energy is converted into particle motion, with a preferred direction determined by the leg inclination~\cite{van2016spatiotemporal,Koumakis2016,baconnier2022selective,caprini2025spontaneous,casiulis2025geometric,delens2026capillary}.

We consider an elliptic particle and tune the shaker amplitude. At large amplitudes, when the shaker is switched on, the particle velocity increases linearly with time along a single trajectory, as expected for Newtonian-like dynamics. In this regime, the particle undergoes accelerated ballistic motion due to effective self-propulsion generated by the directional release of energy during collisions between the legs and the substrate~\cite{scholz2018inertial}.

When the shaker amplitude is reduced, dry friction arising from contact with the substrate can no longer be neglected. At sufficiently low amplitudes, this friction can temporarily halt the motion, effectively pinning the particle. This occurs because the substrate exhibits microscopic random asperities~\cite{Bonn/etal:2014} (Fig.~\ref{fig:1}(c)) that cannot be easily overcome by the small vertical jumps of the particle. These asperities become negligible at high amplitudes, where the vertical excursions of the particle exceed the characteristic scale of the substrate roughness (Fig.~\ref{fig:1}(d)). In this regime, energy dissipation can be effectively described by a Stokes-like friction force proportional to the particle velocity, as reported in previous studies~\cite{caprini2024emergent}. For low vibrational amplitudes, where particles become transiently pinned, we analyze the uninterrupted dynamics after switching on the shaker. Conversely, at high vibrational amplitudes, where pinning is absent, particles are reinitialized in a resting configtion before each measurement.

Figure \ref{fig:2}(a)-(b) shows that, at lower shaker amplitudes, the particle speed increases more sharply with time than at higher amplitudes. In both cases, slow diffusive activations are excluded to distinguish activity-driven motion from passive diffusive dynamics. To characterize the underlying dynamics, we fit the time behavior of the particle speed in the intermediate-velocity regime using the scaling relation $v\sim t^\alpha$. We focus on this regime because, at high velocities, strong vibrations reduce the frequency of activation collisions with the plate, naturally slowing the acceleration, as shown in Fig.~\ref{fig:2}(a)-(b).
These analyses reveal that $\alpha$ approaches the value expected for jerky motion. For the lower shaker amplitude, we obtain $\alpha=1.84\pm0.18$ (Fig.~\ref{fig:2}(a)), consistent with the jerky regime ($\alpha=2$) within experimental uncertainty. By contrast, for the higher shaker amplitude, we find $\alpha=1.28\pm0.03$ (Fig.~\ref{fig:2}(b)), which is closer to the accelerated motion regime ($\alpha=1$). Representative trajectories of both dynamical regimes are shown by thick lines in Fig.~\ref{fig:2}(a)-(b).

{\paragraph{Model} --}
A minimal model to study the two-dimensional active motion displayed by active vibrobots in the low-amplitude shaker's regime is the active Ornstein-Uhlenbeck particle (AOUP) subject to dry friction \cite{antonov2024inertial, antonov2025self}.
Here, the active force is $\mathbf{f}_A = f_A \mathbf{n}$, where $f_A$ represents its amplitude and $\mathbf{n}$ is a time-dependent Ornstein-Uhlenbeck process:
\begin{subequations}
\label{eq:EOM}
\begin{equation}
\label{eq:EOM_a}
    \dot{\mathbf{n}}(t) = -\frac{\mathbf{n}(t)}{\tau} + \sqrt{\frac{2}{\tau}}\boldsymbol{\eta}(t)\,.
\end{equation}
Here $\tau$ is the persistence time, and $\boldsymbol{\eta}(t)$ is a vector of delta-correlated Gaussian white noises. The resulting dynamics for a particle with mass $m$ and velocity $\mathbf{v} = \dot{\mathbf{r}}$ reads:
\begin{equation}
\label{eq:EOM_b}
    m\dot{\mathbf{v}}(t) = -\Delta_F \hat{\mathbf{v}} + \mathbf{f}_A(t) + \sqrt{2K}\,\boldsymbol{\xi}(t)\,,
\end{equation}
where $K$ is the amplitude of the delta-correlated Gaussian white noises $\boldsymbol{\xi}(t)$.
The term $- \Delta_F\hat{\mathbf{v}}$ models dry friction, with $\Delta_F$ denoting the friction threshold value and $\hat{\mathbf{v}}=\mathbf{v}/|\mathbf{v}|$ the unit vector collinear with the velocity.
\end{subequations}
For the vibrobots shown in Fig.~\ref{fig:1}, these white noises are typically small and account for random imperfections in the particle shape and the surface of the plate. 
The vibrobot is pinned at asperities of the rough surface -- i.e., being nearly at rest -- when the magnitude of the active force drops below the dry friction threshold, $|\mathbf{f}_A(t)| \le \Delta_F$, and starts the motion otherwise. This behavior is reflected in the speed probability distribution $p(v),\, v \equiv |\mathbf{v}|$, which is characterized by a peak near zero (Fig.~\ref{fig:2}(c)). The experimental results also demonstrate an overall good agreement with the model. 

{\paragraph{Jerky motion induced by dry friction and activity} --}
To analytically prove that the interplay between activity and dry friction leads to jerky motion, we consider a path-integral approach to understand how a particle initiate its motion starting at rest. 
The problem can be tackled by writing the Onsager-Machlup functional \cite{Caroli1981} for the active particle dynamics starting at rest $\mathbf{n}_0=\mathbf{n}(t_0)=0$ and approaching the final state $\mathbf{n}_f=\mathbf{n}(t_f)$ satisfying $|\mathbf{n}_f| > \Delta_F/f_A$. The latter condition represents a configtion that initiates motion, since the active force is sufficiently high to overcome the dry friction force (see the illustration in Fig.~\ref{fig:3}(a)).

The optimal path minimizing the Onsager–Machlup functional thus determines the most probable activating fluctuation and is obtained from the corresponding Euler–Lagrange equation~\cite{Landau}. Expanding the resulting solution (see End Matter) to first order around the activation time $t_*$, defined by the condition that the active force equals the dry-friction threshold, $\mathbf{n}(t_*) = \hat{\mathbf{v}}\Delta_F/f_A$, yields the short-time solution for $\mathbf{n}(t)$:
\begin{equation}
    \mathbf{n}(t)
    \!=\!
    \frac{\Delta_F}{f_A}\hat{\mathbf{n}}_f
    \left[
    1\!+\!\frac{\cosh(t_*/\tau)}{\tau}(t-t_*)
    \!+\!\mathcal{O}\!\left((t-t_*)^2\right)
    \right]\,.
    \label{eq:activation}
\end{equation}
Substituting this expansion into Eq.~\eqref{eq:EOM} yields the short-time dynamics immediately after the activation,
\begin{subequations}
\begin{eqnarray}
m\dot{\mathbf{v}}(t)&=&-\Delta_F\hat{\mathbf{v}}+f_A\mathbf{n}(t) \label{eq:activ} \\
&\approx&
\frac{\Delta_F}{\tau}(t-t_*)\cosh(t_*/\tau)\hat{\mathbf{n}}_f\,,
\label{eq:v-approx}
\end{eqnarray}
\end{subequations}
where we have neglected orders $(t-t_*)^2$ as well as the contribution of white noise by setting $K=0$. Upon averaging over activation trajectories, the resulting dynamics exhibits a quadratic increase of the speed with time after activation (see Fig.~\ref{fig:3}(b)),
\begin{subequations}
\label{eq:average}
\begin{equation}
    v \approx \frac{\Delta_F\, \mathcal{C}}{2 m \tau}(t-t_*)^2\,,
    \label{eq:quadratic}
\end{equation}
implying a time increase of the displacement as the scaling law:
\begin{equation}
    \frac{m\, \delta r}{\Delta_F\tau^2} = \frac{\mathcal{C}}{6} \left( \frac{t-t_*}{\tau}\right)^3\,,
    \label{eq:scaling}
\end{equation}
where we have introduced the dimensionless constant $\mathcal{C}=\int_0^{\infty} \dd \tilde{t}_*\, p(\tilde{t}_*) \cosh(\tilde{t}_*/\tau)$, with $p(\tilde{t}_*)$ denoting the probability density for the particle to initiate its motion at time $t_* = \tilde{t}_*$. The activation-time distribution is determined by the Onsager-Machlup action (see End Matter).

\begin{figure}[t!]
		\includegraphics[width=\linewidth]{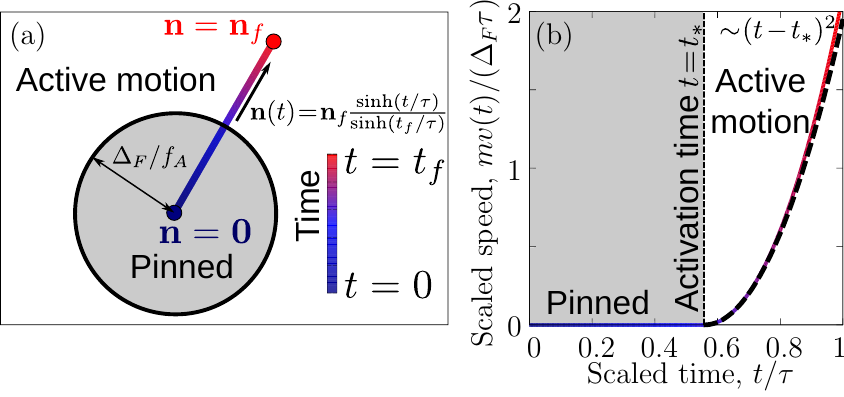}
     \caption{(a) Time color-coded activation path from the pinned state $\mathbf{n}=0$ to the activated state $\mathbf{n}=\mathbf{n}_f$. The approximate form near the activation moment $t=t_*$ (circle boundary) is given by Eq.~\eqref{eq:activation} in the main text, while the full expression is provided in Eq.~\eqref{eq:threshold} of the End Matter. (b) Characteristic speed evolution given by Eq.~\eqref{eq:activ} along this activation path. For $t<t_*$, the particle remains pinned and at rest (gray area), while after the activation at $t=t_*$ the speed exhibits the quadratic growth characteristic of jerky motion. The color coding is identical in the main panel and the inset. The parameters are $t_f=\tau$, $|\mathbf{n}_f|=2$, and $\Delta_F/f_A=1$. 
    \label{fig:3}}
	\end{figure}
    
	 \begin{figure}[t!]
		\includegraphics[width=0.95\linewidth]{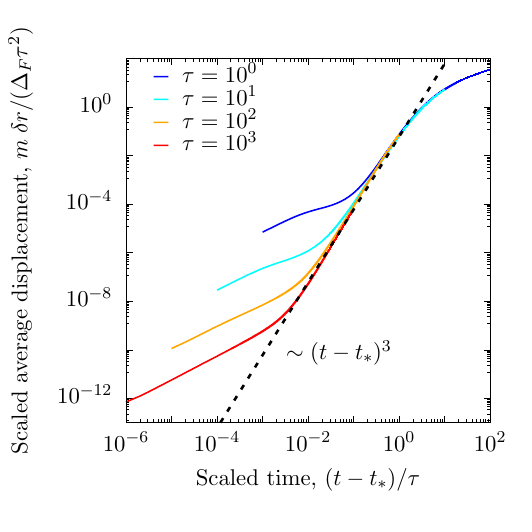}\vspace{-4ex}
     \caption{Scaled average displacement vs scaled time, demonstrating the data collapse onto a common curve. Displacements are calculated numerically from Eq.~\eqref{eq:EOM} with $\Delta_F/f_A = 1$, $K=0$ for the particle initiating its motion after being at rest. The averaging is performed over 5000 such activation events. For $(t-t_*)/\tau \lesssim 1$, the displacement follows the scaling law \eqref{eq:scaling} with $\mathcal{C} = 0.36$, represented by the dashed line.
     }
    \label{fig:4}
	\end{figure}

These scaling laws are consistent with Eq.~\eqref{eq:v_scaling_time} and analytically prove the jerky motion as a consequence of dry friction and activity. Indeed, the displacement dynamics for small time can be equivalently written as
\begin{equation}
    \frac{m \tau}{\mathcal{C}}\dddot{\delta r} = |\mathbf{F}| = \Delta_F = \rm const \,,
\end{equation}
which allows us to find the effective jerk coefficient as $\lambda = m\tau/\mathcal{C}$ by comparison with Eq.~\eqref{eq:jerk}. 
\end{subequations}
The scaling law~\eqref{eq:scaling} is demonstrated in Fig.~\ref{fig:4}, where the simulation data collapse onto a master curve exhibiting the predicted cubic scaling up to the persistence time, $(t-t_*)/\tau \lesssim 1$. The initial ballistic regime reflects slow activation events: particles move briefly following activation but rapidly stop as a result of repinning. For comparison, we applied the same analysis to an equivalent active system with Stokes friction, recovering the conventional scaling laws, $v \sim t$ and $\delta r \sim t^2$ (see End Matter for details). These results demonstrate that the cubic scaling with time of the displacement arises directly from the interplay between activity and dry friction.

From Eq.~\eqref{eq:scaling}, it is evident that the amplitude of the jerky dynamics is enhanced by a short persistence time and a large threshold-to-mass ratio $\Delta_F/m$. The latter is directly related to the dry friction coefficient $\mu$ through $\Delta_F=\mu mg$. However, the conditions that amplify the jerk at the same time also make the detection and observation of the jerky dynamics itself more difficult. Specifically, reducing the persistence time shortens the duration of the jerky motion, which scales with the persistence time (Fig.~\ref{fig:4}), while increasing the dry friction coefficient causes the particle to become more strongly pinned and therefore less frequently activated.

\paragraph{Discussions --}
In this work, we demonstrate through analytical, numerical, and experimental approaches that jerky motion generically emerges from the interplay between activity and dry friction with the substrate. Beyond granular systems~\cite{gnoli2013granular,dauchot2019dynamics,lemaitre2021stress,arbel2025mechanical}, this mechanism is relevant to Brownian motors~\cite{baule2012singular,chen2014large,manacorda2014coulomb,semeraro2023diffusion} and piston-based ratchet systems~\cite{sarracino2013ratchet,sano2014roles}, where nonequilibrium driving and frictional interactions coexist. More broadly, such ingredients are ubiquitous in mechanical systems involving active driving and frictional contacts~\cite{yadav2021stick}, including engines or tribological devices~\cite{persson2013sliding}.

In active granular matter, such jerky motion may provide a mechanism for optimizing the energetic efficiency of self-propulsion, as transient bursts of high jerk can generate rapid changes in motion without requiring constantly large accelerations. Additionally, since jerk is associated with chaotic behavior~\cite{sprott1997some, von1998all, linz2000no, patidar2005bifurcation, njitacke2017antimonotonicity, li2023some}, an interesting open question is how jerk influences collective phenomena, such as flocking~\cite{vicsek1995novel, toner2024physics}, active turbulence~\cite{wensink2012meso, alert2022active}, and motility-induced phase separation~\cite{cates2015motility,bialke2015active}.

\paragraph{Acknowledgments --} A.A. acknowledges funding from the Deutsche Forschungsgemeinschaft (DFG, German Research Foundation) -- Project-ID 570812137. H.L. was supported within the project LO 418/33-1 funded by the Deutsche Forschungsgemeinschaft (DFG).

\paragraph{Data availability --}
The code used for this simulation is publicly available at https://github.com/apantonov/dry-active \cite{Antonov/Lowen:2026}.\ Requests for further information or data should be sent
to the authors.

\section{End Matter}

\begin{figure}[ht!]
		\includegraphics[width=\linewidth]{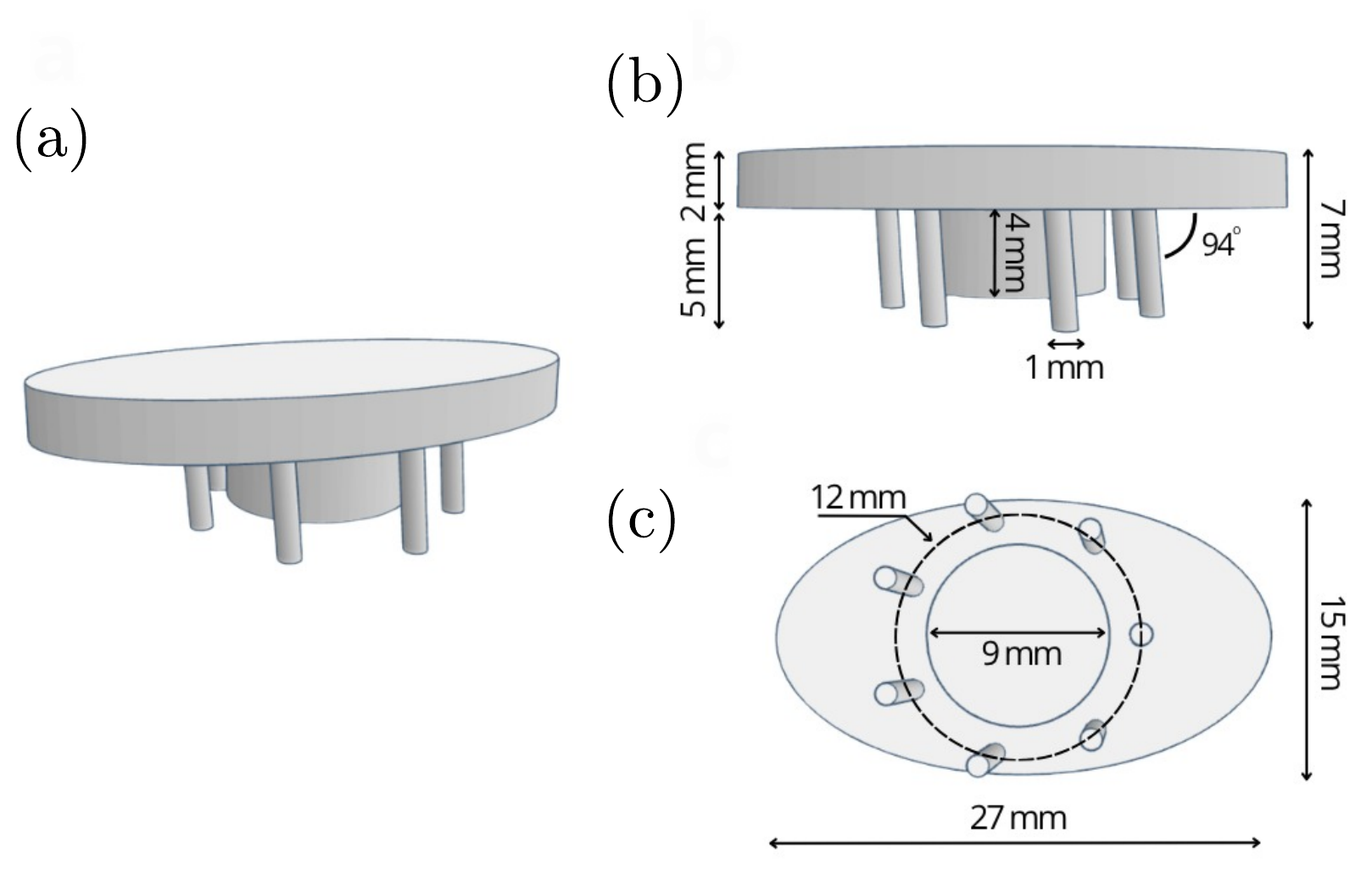}
     \caption{(a) Isometric view of the 3D-printed active granular particle used in the experiments. (b) Its side view, reporting heights of the particle components and the tilting angle of the legs. (c) Bottom view of the particle, showing the diameters of the cylinders forming the particles, the diameter of the legs, as well as the leg positions.}
    \label{fig:5}
	\end{figure}

\paragraph{Particle design}-- 
Active granular particles are manufactured via a proprietary photopolymer using a stereolithographic 3D printer. Each
particle has a body consisting of two concentric cylinders: one elliptical (upper) and one circular (lower). The elliptical cylinder has a height of 2 mm and diameters 15 \si{mm} (smaller diameter) and 27 \si{mm} (larger diameter). The circular cylinder has a height of 4 mm and diameter of 9 mm. Each particle has seven cylindrical legs attached to the elliptical cylinder, with a diameter $1$ \si{mm} and a height of 5 \si{mm}. These legs are tilted in the same direction with an angle of $4^\circ$. Each particle has a mass of $1.45 \pm 0.01$ \si{g}. The design of the particles is provided in Fig.~\ref{fig:5}.

\paragraph{Setup for studying the active motion}-- We place an active granular particle on an acrylic plate whose oscillations are induced by an electromagnetic shaker driven by a conventional function generator. As confirmed in a previous study using a similar setup, the plate oscillations are spatially homogeneous and transfer the same amount of energy to each active granular particle \cite{antonov2025self}. As the shaker amplitude is increased, the particle undergoes vertical jumps with a period controlled by the shaker frequency. The inclined particle legs break the translational symmetry of the particle, leading to asymmetric collisions with the plate. This asymmetry results in directed particle motion when the elastic energy stored during deformation is released~\cite{caprini2024dynamical}. Since the vertical displacement is small compared to the horizontal motion, the dynamics of the active granular particle can be effectively described as quasi-two-dimensional. Finally, imperfections in the plate and particle, combined with residual vertical motion, generate an additional effective translational noise that contributes to the dynamics alongside the active motion~\cite{scholz2018inertial}.
To study active motion in the presence of dry friction, we set the shaker frequency at \SI{75}{\hertz} and the amplitude at $51.186 \pm 0.160$ \si{\micro\meter}, so that the particles are temporarily pinned (i.e. their motion is arrested) during the experiment. To investigate pin-free dynamics, we use the same frequency with an increased amplitude of $65.200 \pm 0.217$ \si{\micro\meter}.

\paragraph{Data acquisition}-- Data are recorded with a high-speed camera placed above the setup, which captures 150 frames per second and is characterized by a spatial resolution of 3.22 \si{px/mm}. Using a tracking algorithm, we extract particle positions, while the orientations are calculated from the relative position of the white spot compared to the particle center of mass. In every recording, images, positions, and orientations are calculated with sub-pixel precision, using conventional image processing techniques. Particles’ velocities $\mathbf{v}$ are calculated by subtracting the position of 16 consecutive frames, $\mathbf{v}(t) = (\mathbf{r}(t+\Delta t/2) -\mathbf{r}(t-\Delta t/2)/\Delta t$, with $\Delta t = 8/75$ \si{sec}. A stopping event is identified when the particle speed $v=|\mathbf{v}|$ satisfies $v \le 1$ \si{mm}. To exclude slow trajectories after such stops, we set the following cut-off conditions: speed $v=30$ \si{mm/sec} needs to be reached within $0.4$ seconds after the activation.

\paragraph{Simulations details}-- Equations~\eqref{eq:EOM} are
numerically solved with the Euler-Maruyama method and a time step $\Delta t = 10^{-3}\tau$. 

\paragraph{Details of analytical calculations}-- {By taking the time derivative of Eq.~\eqref{eq:EOM_b} and eliminating $\dot{\mathbf{n}}$ and $\mathbf{n}$ using Eqs.~\eqref{eq:EOM_a} and~\eqref{eq:EOM_b}, respectively, the dynamics can be recast in the form of Eq.~\eqref{eq:jerk}:}
\begin{align}
   & m\,\tau\ \dddot{\mathbf{r}}(t)  \\
    =&-\left[m + \frac{\Delta_F\, \tau}{|\mathbf{\dot{r}}(t)|}\left(\mathbbm{1}- \hat{\dot{\mathbf{r}}}\hat{\dot{\mathbf{r}}}^T\right) \right]\ddot{\mathbf{r}}(t) -\Delta_F \hat{\dot{\mathbf{r}}}(t) + f_A \sqrt{2\tau}\,\boldsymbol{\eta}(t),\nonumber
\end{align}
which naturally suggests the possibility of jerky dynamics in this model. However, the presence of the third-order derivative alone does not guarantee such behavior, as its contribution may be suppressed by the lower-order terms. Moreover, the emergence of jerky dynamics may depend sensitively on the initial conditions required to initiate the motion.

To address this problem, we pose the following question: how does a particle initiate its motion after being at rest? Within the proposed model, this question can be reformulated in terms of the path integral formalism:
\begin{subequations}
    \begin{equation}
    \label{eq:PI}
    \mathcal{P}(\mathbf{n}_f,t_f|\mathbf{n}_0, 0) = \int\limits_{\mathbf{n}(0) = \mathbf{n}_0}^{\mathbf{n}(t_f) = \mathbf{n}_f} \mathcal{D}\mathbf{n}(t) \exp\left(-\frac{S[\mathbf{n}(t)]}{\tau} \right),
\end{equation}
where $\mathcal{P}$ is the probability to reach the running state $\mathbf{n}_f$, starting from the state at rest $\mathbf{n}_0=0$, within time $t_f$; and
\begin{equation}
    S[\mathbf{n}(t)] = \frac{1}{4}\int\limits_0^{t_f} \dd t \left|\dot{\mathbf{n}} + \frac{\mathbf{n}}{\tau}\right|^2 - \frac{t_f}{\tau^2}
    \label{eq:OM}
\end{equation}
is the corresponding Onsager-Machlup functional~\cite{Caroli1981}. 
\end{subequations}
The optimal path minimizing the Onsager–Machlup functional thus determines the most probable activating fluctuation and is obtained from the corresponding Euler–Lagrange equation~\cite{Landau}:
\begin{subequations}
    \begin{align}
        \frac{\dd}{\dd t}\frac{\partial \mathcal{L}}{\partial{\mathbf{\dot n}}} = \frac{\partial \mathcal{L}}{\partial{\mathbf{n}}}\,,
    \end{align}
    with the Lagrangian
    \begin{equation}
        \mathcal{L}(\mathbf{n}, \mathbf{\dot n}, t) = \frac{1}{4}\left|\dot{\mathbf{n}} + \frac{\mathbf{n}}{\tau}\right|^2\,.
    \end{equation}
\end{subequations}
This yields the second-order differential equation
\begin{equation}
    \ddot{\mathbf{n}}(t) = \frac{\mathbf{n}(t)}{\tau^2}\,,
\end{equation}
which for the boundary conditions in Eq.~\eqref{eq:PI} has the solution
\begin{equation}
    \mathbf{n}(t) = \mathbf{n_f}\frac{\sinh(t/\tau)}{\sinh(t_f/\tau)} \,.
    \label{eq:threshold}
\end{equation}
Expanding Eq.~\eqref{eq:threshold} to first order around the activation time $t_*$, namely the time where the active force equals the dry friction one, such that $\mathbf{n}(t_*) = \hat{\mathbf{v}}\Delta_F/f_A$ allows us to find a solution for $\mathbf{n}(t)$ valid for small times, which gives Eq.~\eqref{eq:activation} from the main text.

For the ensemble average performed in Eqs.~\eqref{eq:average}, we integrate over the activation time $\tilde{t}_*$, which parametrizes the center of the activation trajectories in Eq.~\eqref{eq:activation} \cite{coleman1988aspects}. The corresponding probability density of the activation time is related to the Onsager-Machlup action~\eqref{eq:OM} according to
\begin{equation}
    p(\tilde{t}_*) = A(\tilde{t}_*) e^{-S[\mathbf{n}(t|\tilde{t}_*)]}\,,
\end{equation}
where $A(\tilde{t}_*)$ accounts for stochastic fluctuations around the most probable activation path (Eq.~\eqref{eq:activation}) associated with the activation time $\tilde{t}_*$~\cite{Caroli1981, coleman1988aspects}, which we denote as $\mathbf{n}(t\mid\tilde{t}_*)$. The ensemble-averaged speed therefore reads:
\begin{align}
    v &= \int_0^{\infty} \dd \tilde{t}_*\, |\mathbf{v}[\mathbf{n}(t\mid\tilde{t}_*)]|\, p(\tilde{t}_*)\nonumber\\
    &\approx \int_0^{\infty} \dd \tilde{t}_*\, |\mathbf{v}[\mathbf{n}(t- t_*)]|\, p(\tilde{t}_*)\,,
    \label{eq:av-approx}
\end{align}
where $\mathbf{v}[\mathbf{n}(t\mid\tilde{t}_*)]$ is the velocity conditioned by the activation path $\mathbf{n}(t\mid\tilde{t}_*)$. In Eq.~\eqref{eq:av-approx}, we further approximate it to depend only on the time since activation, $t-t_*$. Integrating $\mathbf{v}[\mathbf{n}(t- t_*)]$ given by Eq.~\eqref{eq:v-approx} once with respect to $t-t_*$ and substituting the result into \eqref{eq:av-approx} yields Eq.~\eqref{eq:quadratic} from the main text.
The displacement averaging is performed analogously. 
In Eq.~\eqref{eq:av-approx}, we assume that the velocity after activation at time $t_*$ is independent of the activation point itself, such that $\mathbf{v}[\mathbf{n}(t\mid\tilde{t}_*) \approx \mathbf{v}[\mathbf{n}(t- t_*)]$, thus neglecting the contribution of fluctuations around the escape path, encoded in the prefactor $A(\tilde{t}_*)$, to the velocity. This approximation is well justified in the activation regime, where the dynamics is dominated by the most probable activation trajectory. Moreover, the numerical results in Fig.~\ref{fig:4} indicate that the resulting quantity $\mathcal{C}$ is approximately independent of persistence time $\tau$. However, the situation is less straightforward during the initial onset of motion. In particular, inertial memory combined with the discontinuity of the dry friction force at zero velocity may couple fluctuations in the activation trajectory to the subsequent velocity. This coupling is present, e.g., in the initial ballistic regime shown in Fig.~\ref{fig:4}.

{\paragraph{Speed and displacement for an active particle with Stokes friction}--} We note that the presence of pinning by the dry friction is crucial for emergence of the jerky motion. To demonstrate this, we replace the frictional force in Eq.~\eqref{eq:EOM}, used in the analysis in the main text, by a linear Stokes drag,
\begin{equation}
    m\dot{\mathbf{v}}(t)
    =
    -\gamma\,\mathbf{v}(t)
    +f_A\mathbf{n}(t),
\end{equation}
where $\gamma$ is the viscous Stokes friction. At the same time, we retain the same activation state \eqref{eq:activation} as the initial condition. In this case, the first-order expansion after activation results in accelerated motion instead.
The time behavior of the speed $v$ reads
\begin{subequations}
\begin{align}
    v &= 
\frac{\Delta_F}{\gamma \tau}\Bigg[ e^{-\frac{\gamma(t- t_*)}{m}} \left(\frac{m}{\gamma} \cosh \left(\frac{t_*}{\tau}\right)- \tau\right)\nonumber \\
&+\tau-\cosh \left(\frac{t_*}{\tau}\right) \left(t_*-t+\frac{m}{\gamma}\right)\Bigg]+\!\mathcal{O}(t-t_*)^2\nonumber \\
& = \frac{\Delta_F}{m}(t-t_*) + \mathcal{O}(t-t_*)^2\,,
\end{align}
while the displacement $\delta r$ as a function of time has the following expression:
\begin{align}
    \delta r&
    =
    \frac{\Delta_F}{2m}(t-t_*)^2
    +\mathcal{O}(t-t_*)^3\,,
\end{align}
\end{subequations}
consistent with the scaling of a Newtonian dynamics.

\end{document}